\documentclass[
	a4paper,
	10pt,
	unnumberedsections, 
	twoside, 
]{LTJournalArticle}
\usepackage[dvipsnames]{xcolor}
\usepackage{lineno,hyperref,xcolor,float,caption,paralist,siunitx,amssymb} 
\usepackage[singlespacing]{setspace}
\usepackage{soul} 
\modulolinenumbers[5]
\usepackage[T1]{fontenc}
\usepackage{wrapfig}
\usepackage{multicol}
\usepackage{changes}

\usepackage{amssymb}
\usepackage{amsthm}
\usepackage{graphicx} 
\usepackage{amsmath}
\usepackage{upgreek}
\usepackage{float}
\usepackage{xcolor}
\usepackage{subcaption}
\usepackage{siunitx}
\usepackage{gensymb}
\usepackage[T1]{fontenc}
\usepackage{placeins}
\usepackage{ulem}

\title{Design and characterisation of an antiproton deceleration beamline for the PUMA experiment}

\author{
	J. Fischer\textsuperscript{a}\thanks{Corresponding author: \href{mailto:jfischer@ikp.tu-darmstadt.de}{jfischer@ikp.tu-darmstadt.de}}, 
 A. Schmidt\textsuperscript{a},
 N. Azaryan\textsuperscript{b},
 F. Butin\textsuperscript{b},
 J. Ferreira Somoza\textsuperscript{b},
 A. Husson\textsuperscript{a},\\
 C. Klink\textsuperscript{a}, 
 A. Obertelli\textsuperscript{a}, 
 M. Schlaich\textsuperscript{a},
 A. Sinturel\textsuperscript{b},
 N. Thaus\textsuperscript{b},
 and F. Wienholtz\textsuperscript{a}
}

\date{\footnotesize\textsuperscript{\textbf{a}}Technische Universität Darmstadt, Institut für Kernphysik, Schloßgartenstraße 9, 64289 Darmstadt, Germany\\ \textsuperscript{\textbf{b}}CERN, 1211 Geneva 23, Switzerland}

\renewcommand{\maketitlehookd}{
	\begin{abstract}
		\noindent We report on the design and characterization of an antiproton deceleration beamline, based on a pulsed drift tube, for the PUMA experiment at the Antimatter Factory at CERN. 
The design has been tailored to high-voltage ($100\,\si{kV}$) and ultra-high vacuum (below $10^{-10}$\,\si{mbar}) conditions.
A first operation achieved decelerating antiprotons from an initial energy of $100\,\si{keV}$ down to $(3898 \pm 3)$\,\si{eV}, marking the initial stage in trapping antiprotons for the PUMA experiment. 
Employing a high-voltage ramping scheme, the pressure remains below $2 \times 10^{-10}$\,\si{mbar} upstream of the pulsed drift tube for $75\%$ of the cycle time.
The beamline reached a transmission of $(55 \pm 3)$\% for antiprotons decelerated to 4\,\si{keV}.
The beam is focused on a position sensitive detector to a spot with horizontal and vertical standard deviations of $\sigma_\mathrm{horiz}=(3.0 \pm 0.1)$\,\si{mm} and $\sigma_\mathrm{vert}=(3.8\pm 0.2)$\,\si{mm}, respectively.
This spot size is within the acceptance of the PUMA Penning trap. 
	\end{abstract}
}

\begin{document}
\maketitle 

\section{Introduction}
\label{sec:introduction}
The spatial distribution of protons and neutrons at and beyond the nuclear surface of atomic nuclei challenges nuclear theory.  In particular, nuclei with a neutron excess exhibit a so-called neutron skin, where the neutron density distribution extends beyond the proton density distribution. 
The thickness of the neutron skin is defined as the difference in root-mean-square radii of the density distributions
\begin{align}
\Delta r_\mathrm{np} = \langle r^2_\mathrm{n} \rangle^{1/2} -  \langle r^2_\mathrm{p} \rangle^{1/2}.
\end{align}
The neutron skin thickness correlates with the slope parameter $L$ of the nuclear equation of state \cite{Roca-Maza11}, playing an important role in defining the relation between the mass and radius of a neutron star \cite{Hebeler13,Horowitz01}. 
Neutron skin thicknesses have been investigated with several methods \cite{KRASZNAHORKAY04,Klos07,Zenihiro10,Tarbert14,Adhikari21}, mostly on stable nuclei, while the challenge lies in determining the radius of the neutron distribution $\langle r^2_\mathrm{n} \rangle^{1/2}$ with enough accuracy and controlled theoretical uncertainties. 
Information on unstable nuclei is much more scarce, as illustrated by Ca isotopes: charge radii can be accessed with precision from the relative measurement of isotope shifts from laser spectroscopy and anchored to stable nuclei \cite{Ruiz2016}, while the interpretation of the data related to the matter or neutron radius suffers from model dependence \cite{Wakasa2023, Enciu2022}. 
Nuclei close or at the neutron drip line can have loosely bound nucleons, whose wave function extends far beyond the charge distribution. 
Such systems are called halo nuclei \cite{Tanihata85,Hansen95}. 
Neutron halos have been so far observed in light nuclei only \cite{JONSON2004}.
Indications for $p$-wave halos in medium mass nuclei have been reported \cite{Manju2019}, while more halos are predicted to exist in uncharted regions of the nuclear landscape \cite{Rotival2009}.
Proton halos have been predicted as well \cite{Ren1996}. \\
Most aforementioned methods to probe neutron skins and halos in stable and unstable nuclei are sensitive to the nuclear surface where $\rho\sim\rho_0/2$, not further out in the tail of the density distribution, where the asymmetry is the largest.
The antiProton Unstable Matter Annihilation (PUMA) experiment aims to investigate these phenomena in the tail of stable and unstable nuclei with low-energy antiprotons as a probe \cite{WADA2004,Puma22}. 
Antiprotons are uniquely suited for this, as they annihilate with nucleons at a mean radial position $\sim 2$\,\si{fm} further out from the half density radius of the nucleus \cite{Klos07,LEON74,ILJINOV82}, probing a region of higher neutron-to-proton asymmetry. 
The PUMA experiment will produce antiprotonic atoms by combining nuclei and antiprotons in a Penning trap. 
By studying the pions produced in the annihilation, the PUMA experiment can determine the neutron-to-proton ratio in the tail of the nuclear density distribution. 
The setup is located at the Antimatter Factory at CERN. 
Stable isotopes are supplied by an offline ion source \cite{Schlaich23}, and for the investigation of more neutron-rich and unstable isotopes the setup will be transported to the ISOLDE facility \cite{Borge2018} at CERN.\\
The ELENA ring at the Antimatter Factory provides bunches of $5\cdot 10^6$ to $10^7$ antiprotons at 100\,\si{keV} to up to four experiments every 2 minutes \cite{Bartmann2018,Maury14,Ponce22}. 
To further decelerate the antiprotons to energies compatible with the PUMA Penning trap, one can use a thin degrader foil or pulsed drift tubes (PDT) \cite{Kalinowsky1993,AMOLE2014,Tajima2019,HUSSON21,Amsler2021,Latacz2023}. 
Employing a foil for deceleration is space-efficient, but the yield is low and the energy distribution broad \cite{Nordlund2022}, compared to a pulsed drift tube, which can have a transmission of 100\% while conserving the width of the energy distribution. 
For antiprotons with an initial energy of approximately 100 keV, trapping efficiencies vary from a few percent \cite{Kuroda05} to a maximum of 50\%, predicted in \cite{fabbri19}. 
However, for the PUMA experiment, which relies on the simultaneous trapping of antiprotons and stable and unstable ions, the use of a foil is unfeasible, since low-energy ions cannot penetrate the foil.\\
An established method to change the energy of a particle beam is to use a drift tube, where the potential can be changed rapidly.
Here, the drift tube is set to a potential and is used to decelerate the particles to the desired energy. 
If the electrode is switched to a different potential, \textit{e.g.}, ground, while the particles are still inside and in the field free region of the drift tube, they are not reaccelerated on exit. 
Because only the longitudinal and not the transversal kinetic energy is changed, the divergence angle of the beam increases by a factor of $ \sqrt{E_\mathrm{in}/E_\mathrm{out}}$, where $E$ is the kinetic energy of the incoming and outgoing particles, respectively.
This can be compensated by additional ion optical elements or beam cooling.\\
Several ion trap experiments use pulsed drift tubes to decelerate nuclei for trapping \cite{Schlaich23,HERFURTH01,COECK07,GRUND20}, often in combination with buffer-gas cooling \cite{SAVARD1991} to counteract the increase in transversal emittance, some from energies as high as $60$\,\si{keV}. 
The GBAR experiment at CERN is confronted with a similar problem as the PUMA experiment, as they need to decelerate antiprotons to $1$\,\si{keV} \cite{Perez15}. 
At the PUMA experiment, the antiprotons are decelerated from 100\,\si{keV} to 4\,\si{keV} to allow for an efficient beam transport and in a second step down to 100\,\si{eV} right in front of the trap. \\
To limit the annihilation of antiprotons with residual gas molecules, a vacuum of a few $10^{-10}\,\si{mbar}$ along and $10^{-11}\,\si{mbar}$ at the end of the beamline is critical.

\section{Beamline Design}\label{sec:beam_line_design}
\subsection{Transfer Line from ELENA to PUMA}
\begin{figure}[h]
    \centering
        \includegraphics[width=\linewidth]{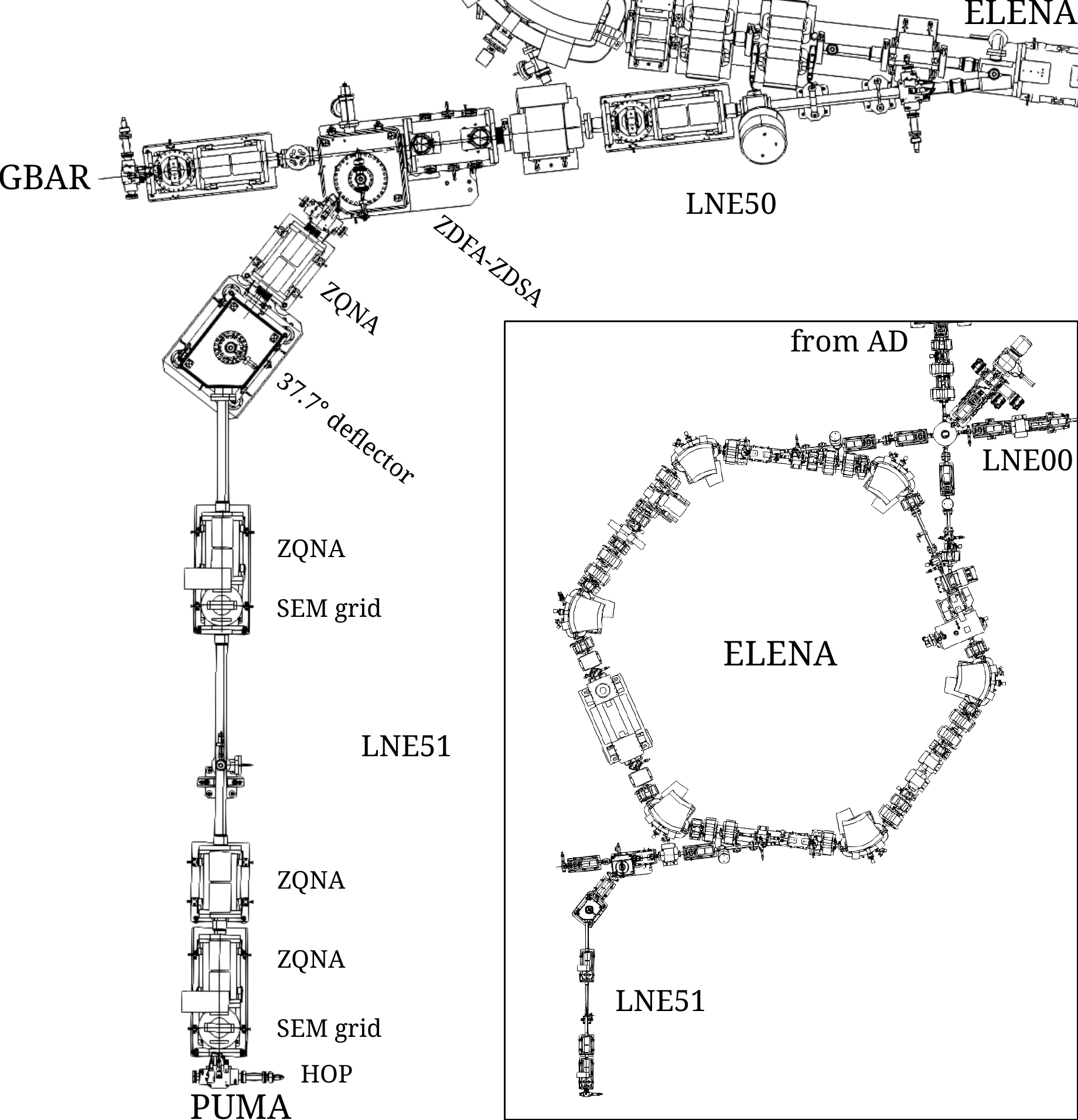}
        \caption{Schematic view of LNE51 transfer line to PUMA. Antiprotons are ejected from ELENA into LNE50, from which LNE51 branches off. The insert shows the position of LNE51 relative to ELENA and LNE00.}
        \label{Fig:LNE51}
\end{figure}
The transfer of 100\,keV particles (H$^-$ ions or antiprotons) from the ELENA machine to the PUMA experiment is performed by the so-called LNE51 transfer line. 
LNE51 branches off from the LNE50 line (transfer from ELENA to the adjacent GBAR experiment) using a standard ZDFA-ZDSA switching unit (fast switch and electrostatic deflector) integrated in LNE50. 
This equipment is interlocked with the access safety system of the PUMA zone, preventing any beam to be sent from the ELENA machine, while the area is being accessed.
The sector valve at the interface between the experiment and the LNE51 transfer line is interlocked with the access system to close automatically when the zone is being accessed. 
This drastically limits the risk of contamination of the upstream sections of ELENA machine in case of an incident while manipulating the experimental equipment.
To satisfy the integration constraints and match the beam to the PUMA experiment at the end of the line, four electrostatic quadrupole/H-V corrector units (ZQNA) are installed, along with a 37.7\degree\,standalone deflector. 
At the focal point, the beam spot size (rms) is approximately 2\,mm and the horizontal and vertical geometric emittance ($95\% = 6\epsilon_\mathrm{rms}$) is 6\,mm\,mrad and 4\,mm\,mrad, respectively \cite{Fraser15}.
The layout for LNE51 is shown schematically in Figure\,\ref{Fig:LNE51}. 
Two SEM grids (Secondary Emission Monitors) \cite{Martini1997} are installed in LNE51.
They are standard equipment in the ELENA transfer lines that allow to extract the profile of the impinging beam, either H$^-$ ions or antiprotons. 
Made from x-y meshes of 50\,$\upmu$m tungsten wires, covering the beam acceptance, spaced by a pitch of 0.5\,mm in the central region, they intercept only about 10\% of the beam at each station \cite{mclean2021}. 
These monitors are ultra-high vacuum compatible, as they can be baked-out to 200\degree C.
As bake-out is required, the vacuum line is fitted with permanently installed bake-out jackets.

\begin{figure*}[h]
	\centering
	\includegraphics[width = \textwidth]{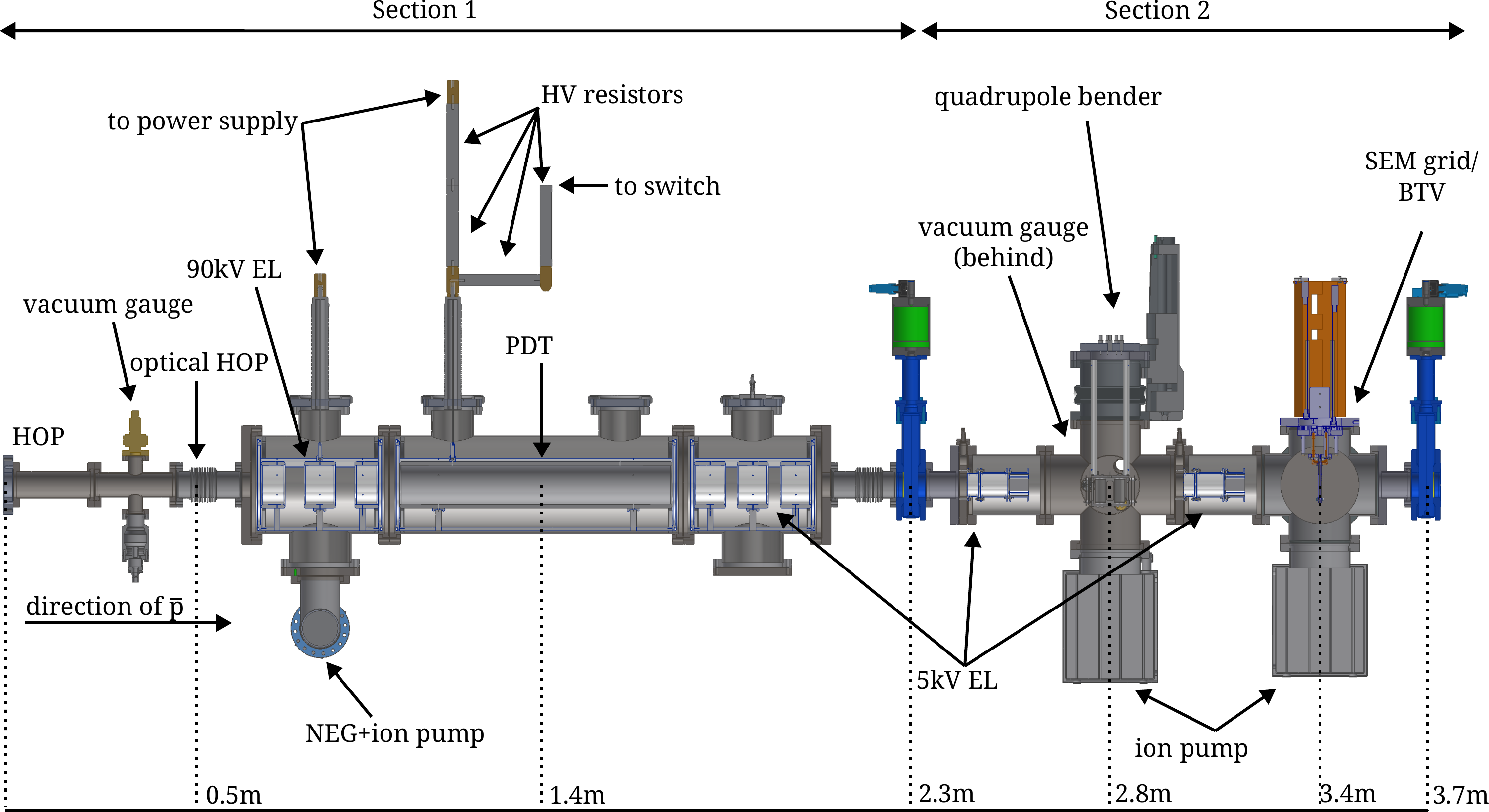}
	\caption{Half-section view of the beamline without the supports. The antiprotons traverse the beamline from left to right. Ions from the offline source enter the beamline at the quadrupole bender in the direction into the page and are deflected to the right. The gate valves separating the sections are depicted in blue.}
	\label{Fig:setup}
\end{figure*}

\subsection{The PUMA Antiproton Beamline}
Downstream of the handover point (HOP) to PUMA (see Fig.\,\ref{Fig:LNE51} and \ref{Fig:setup}), the beamline consists of two main sections, that can be isolated by gate valves type 48236-CE44 from VAT (see Fig.\,\ref{Fig:setup}). 
Section 1 includes the pulsed drift tube itself. 
It is complemented by a high-voltage (up to -85\,kV) as well as a low-voltage (up to 5\,kV) einzel lens (EL) on the injection and ejection sides, respectively, to focus the antiproton bunches into and out of the pulsed drift tube.\\
Section 2 consists of two low-voltage (up to 5\,kV) einzel lenses with x-y-steerers to guide the beam to the entrance of the PUMA Penning trap. 
In between these lenses, a quadrupole ion beam bender allows the injection of ions from an offline ion source setup, perpendicular to the antiproton beamline. 
Even tough the bender has been designed to allow for simultaneous injection of ions and antiprotons, it can be removed when it is not needed.  
A beam imaging system (BTV), which consists of a phosphorous screen and a camera, completes the section. 
The BTV can be moved in and out of the beamline, as it is a completely destructive measurement of the beam.
In the future, the BTV will be replaced by a SEM grid. 

\begin{figure}[h]
    \centering
        \includegraphics[width=\linewidth]{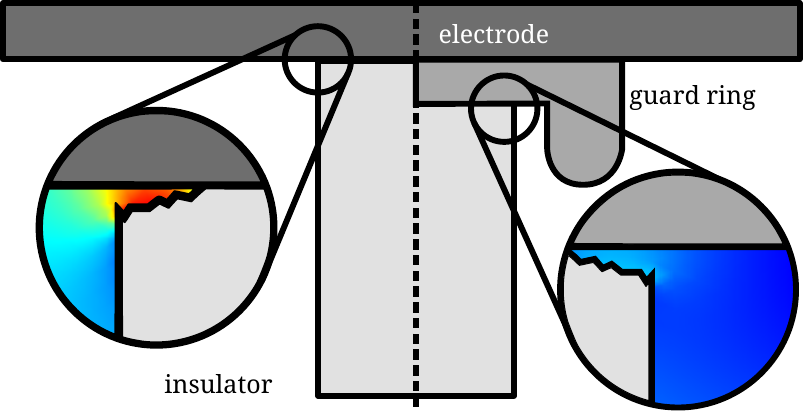}
        \caption{The field strength at an unshielded triple junction (left) and one shielded with a guard ring (right) is illustrated here. Blue indicates lower and red higher electric field strengths.}
        \label{Fig:guardring}
\end{figure}

\subsection{The Pulsed Drift Tube}\label{subsec:PDT}
The pulsed drift tube (PDT) used for the PUMA experiment, is based on the GBAR design \cite{HUSSON21}. 
Although the high-voltage einzel lens in front of the drift tube counteracts the strong focusing effect of the decelerating electric field, the drift tube has to accommodate an expansion of the beam. 
The inner diameter was thus chosen to be 100\,\si{mm} with an outer diameter of 120\,\si{mm}. 
At 4\,\si{keV}, an antiproton bunch from ELENA has a length of 250\,\si{mm} ($2\sigma$) \cite{Ponce22}. 
The PUMA pulsed drift tube has been designed to be 700\,\si{mm} long. 
This ensures, that the bunch is in the field free region of the drift tube when the potential is changed. \\
Because of the stringent vacuum requirements ($p<10^{-10}\,\si{mbar}$), materials with the lowest possible outgassing rates have to be used.
Therefore, the pulsed drift tube is made from aluminium ($\sim 1\cdot 10^{-13}\,\si{mbar~l/s/cm^2}$), which outgases less than stainless steel ($\sim 3\cdot 10^{-12}\,\si{mbar~l/s/cm^2}$) \cite{Benvenuti1988}. 
The insulators are made from MACOR\textsuperscript{\textregistered}, which has an outgassing rate of $1.1\cdot 10^{-11}\,\si{mbar~l/s/cm^2}$ \cite{Battes2021}.\\
The walls of the vacuum chambers are coated with a non-evaporable getter (NEG) to pump the section.
Non-evaporable getters are made from an alloy of Zr, V, Ti, Al and Fe, that can sputtered directly onto the wall of a vacuum chamber \cite{CHIGGIATO06}. 
It acts as a pump by absorbing hydrogen and chemically binding other reactive gases like oxygen.
To activate the NEG, the chambers are heated (200\degree C to 400\degree C).
Molecules at the surface (mainly carbon, nitrogen and oxygen) diffuse into the bulk.
Hydrogen is released and must be pumped away by another pump.
Therefore, all components, such as vacuum gauges, valves, feedthroughs, pumps, cables and beam instrumentation, must be bakable at 250\degree C at least.
The coating of the inside surfaces of the chambers was done at CERN.
The installation of the pulsed drift tube inside the chamber must be done without touching the coating to prevent damaging it.
It is first mounted onto its support structure before being lowered vertically into the vacuum chamber and secured with screws.
To facilitate individual access to the high- and low-voltage einzel lens as well as the drift tube, the vacuum chamber is divided into three parts.\\
At the intersections of vacuum, conductor and insulator, the electric field is strongly enhanced due to gaps arising from imperfections on the corners of the material (see Fig.\,\ref{Fig:guardring}). 
Special attention has been paid to these so-called triple junctions to prevent possible discharges. 
They are shielded by purpose-built rings, that surround the triple junction and thereby lower the electric field (see Fig.\,\ref{Fig:guardring}). 
On all components, sharp edges have been avoided, and the electrodes have been polished to an average surface finish of $R_\mathrm{a}=0.05\,\si{\upmu m}$, which helps to prevent discharges \cite{Williams1972}.

\subsubsection{Electronics}\label{subsec:electronics}
To not reaccelerate the antiprotons as they exit the pulsed drift tube, it must be discharged from -96\,kV to 0\,V before the first antiprotons exit the field free region of the drift tube. 
For antiprotons with a kinetic energy of 4\,keV, the time to discharge the drift tube is in the order of 500\,ns.
Equipment that can withstand high voltages and high peak currents, as well as a high-voltage switch with a short transient, are needed.
The pulsed drift tube is connected to a high-voltage power supply (Spellman~SL130PN60) via a $\SI{1}{\mega\ohm}$ resistor. 
In order not to exceed the voltage rating of the resistors, two Metallux~HVR~969 resistors are used, connected via polished brass cylinders with rounded edges.
The value is chosen as a compromise between the need for a high resistance to decouple the power supply from the pulsed drift tube while switching, and the need for a low resistance to minimize the effects of current fluctuations on the voltage applied to the pulsed drift tube. 
For the discharge of the tube's capacitance, a fast high-voltage switch (Behlke HTS 1501-20-LC2) connects the pulsed drift tube to ground. 
To make sure that the switch is not damaged, the pulsed drift tube is connected to the switch via a two $\SI{250}{\ohm}$ Metallux~HVR~969 resistors in series, limiting the current.
The high-voltage leads are connected with HN-70 connectors from R.E.~Beverly~III~\&~Associates. 
The cables are suspended from the ceiling to avoid triple junctions at the exposed high-voltage connectors. 
The grounded mesh is removed on the load side, and special care is taken to cover the pointy ends of the grounded mesh. 
As high-voltage feedthrough, a HV125R-CE-CU39 from VACOM, rated for up to 125\,kV is used.\\

\begin{figure}
	\centering
	\includegraphics[width = 1.0\linewidth]{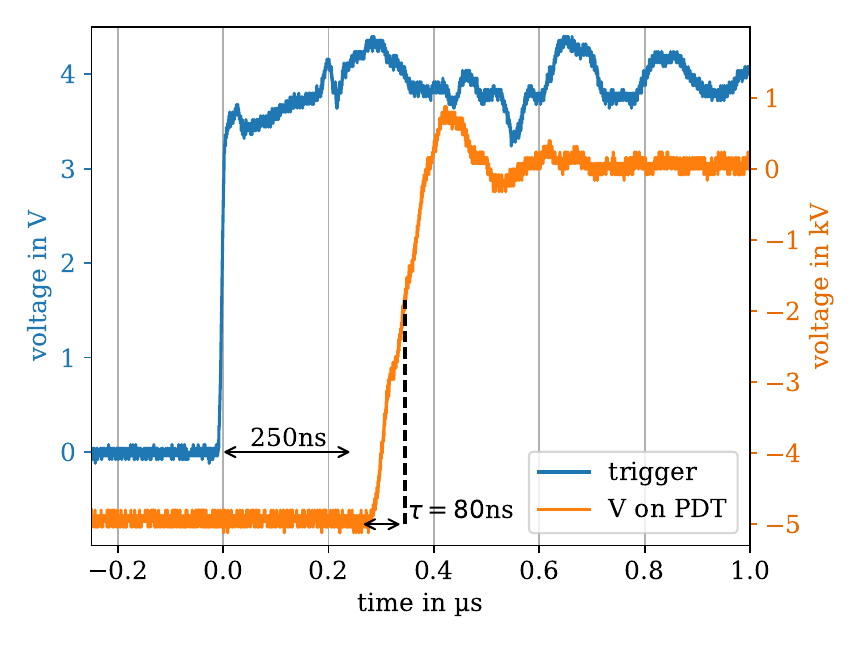}
	\caption{Switching time while switching from $5\,\si{kV}$ to ground, measured with a $1/1000$ voltage divider. The trigger signal is shown in blue and the voltage on the pulsed drift tube in orange.}
	\label{fig:PDT switching time}
\end{figure}

Using a 1/1000 voltage divider (LeCroy PPE6kV) connected to a Tektronix MDO3104 oscilloscope, the switching time from -5\,kV to ground was measured.
As can be seen in Fig.\,\ref{fig:PDT switching time}, there is a $\sim$250\,ns delay between the trigger signal (blue) and the voltage on the pulsed drift tube (orange) which has to be taken into account when triggering the switch. 
Independent of the voltage applied to the switch, the transient time $\tau$ to $V_0/\mathrm{e}$ is $\sim$80\,ns which is consistent with the time constant estimated by a simple RC-circuit, where the capacitance of the pulsed drift tube was measured and within the specs of the switch: 
\begin{align}
    \tau=RC=500\Upomega\cdot 170\mathrm{pF}=85 \, \mathrm{ns}.
\end{align}

\subsubsection{Safety Cage}
The high-voltage system has unshielded $\sim$ 100\,kV connections exposed to air during operation. 
Therefore, the safety of the users has to be ensured by a safety cage according to the ingress protection code level IP3X.
Following the European norm EN 50191, the dimensions of the safety cage are defined so that any high-voltage point in air is at a distance of more than 74\,\si{cm} from the cage, corresponding to a maximum voltage of 130\,kV, the maximum voltage of the high-voltage power supply. 
The high-voltage system is interlocked via a switch (Telemecanique XCSDMC7902 coded magnetic switch) at the sliding door of the cage to interlock the power supplies in the event of unexpected access while the equipment is powered.
The safety cage is further secured with a trapped key system from Allen Bradley (Rockwell) to prevent unauthorized access. 
It must first be locked to be able to switch on the high-voltage power supplies.
To simplify maintenance work, panels can be removed from all sides of the cage.

\section{Vacuum and Conditioning}
\subsection{Baseline Vacuum Pressure}
Due to the strict vacuum requirements at the entrance of the PUMA trap, special attention must be paid to the pressure. 
After activating the NEG coating, a pressure of $2\cdot 10^{-11}$\,mbar was measured at the end of the pulsed drift tube section, a factor of 10 better than required. 
For the subsequent tests, the NEG coating was not reactivated after venting, to conserve it for the use with the PUMA trap attached.
Without the NEG activated, the pressure base level is around $1.4\cdot 10^{-10}$\,mbar. 
This is sufficient to condition and operate the pulsed drift tube.

\subsection{High-Voltage Conditioning}
Surface contamination and imperfections are sources of discharges that degrade the vacuum and material when high voltage is applied. 
They also lead to a leakage current that drains the set potential. 
This difficulty can be countered by conditioning the high-voltage parts, which is therefore an essential step before operating the pulsed drift tube.
It was done by a stepwise increase of the voltage, while keeping the leakage current below the limit of the power supply and the vacuum better than $5\cdot 10^{-8}$\,mbar.

The pulsed drift tube and high-voltage einzel lens were conditioned over several weeks. 
The voltage was increased step by step and left in static operation until the sudden spikes in current, associated with field emission from imperfections on the electrode, subsided, which took between 12 and 72 hours per voltage step. 
In addition to the conditioning, modifications to the setup were made outside the vacuum to reduce the leakage current.
These focussed on increasing the distance from any high-voltage parts to ground, as well as polishing and rounding pieces in high electric fields.
Ultimately, the leakage current at -96\,kV could be lowered from 100\,$\upmu$A to 50\,$\upmu$A by polishing and increasing the corner radius of one high-voltage part from 3\,mm to 15\,mm. 
Additionally, the current could be further decreased to 11\,$\upmu$A by increasing the ceiling height of the safety cage by 50\,cm to 75\,cm.
The leakage current of the high-voltage einzel lens could not be reduced in the same way. 
At -85\,kV, the 100\,$\upmu$A current limit of the power supply is reached. 
This means that the design value of -90\,kV could not be achieved, nevertheless it could be used for commissioning.
A redesign with increased distances between high-voltage parts and ground is planned.

\subsection{Vacuum During Operation}
\begin{figure}[ht]
	\centering
	\includegraphics[width = \linewidth]{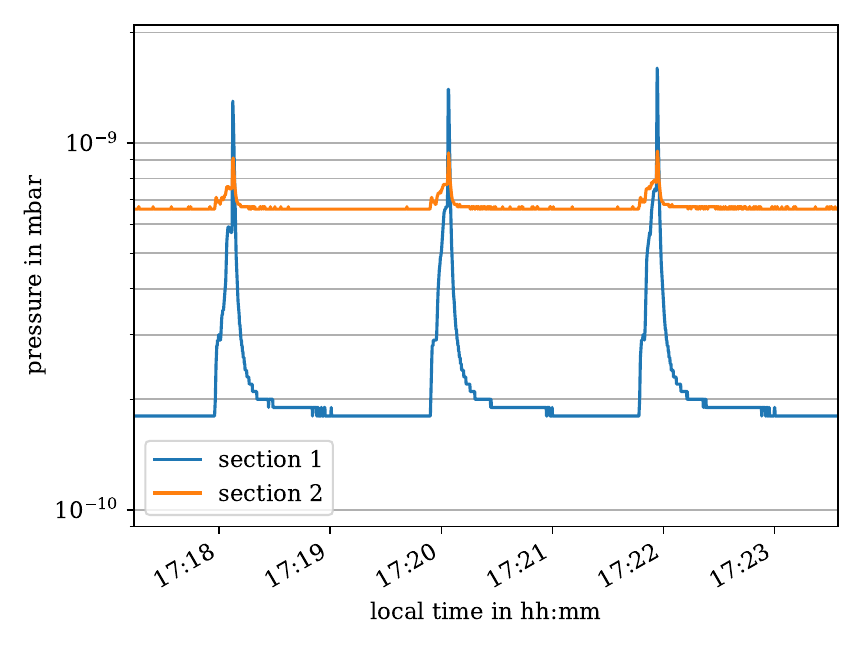}
	\caption{The pressure in section 1 and in section 2, while switching (three cycles). This is without the NEG coating activated, to conserve it for the use with the PUMA trap attached.} 
	\label{Fig:pressure-while-switching}
\end{figure}

During operation of the pulsed drift tube, the remaining leakage current inside the vacuum degrades the pressure. 
To mitigate this, as done by the GBAR collaboration,  the voltage is kept at 0\,V for most of the ELENA cycle and is increased to -96\,kV only 9.5\,s before a bunch of antiprotons arrives. 
Ramping up the voltage only shortly\footnote{compared to a repetition time of 120\,s for ELENA.} before the bunch arrives has the advantage, that the vacuum is below $2\cdot 10^{-10}$\,mbar most of the time, since there is no leakage current at 0\,V. 
When -96\,kV are applied, the pressure reaches a value of $8\cdot 10^{-10}$\,mbar and increases to $2\cdot 10^{-9}$\,mbar when switching (see Fig.\,\ref{Fig:pressure-while-switching}). 

\section{Measurement of Beam Properties}\label{sec:beam_properties}
\subsection{Detection System}\label{subsec:detection}
For the characterization of the system, a vacuum chamber with several detectors was installed at the end of the beamline. 
To visualize the beam spot, a microchannel plate (MCP) by Hamamatsu with a phosphor screen with a diameter of 40\,mm was used. 
In combination with the camera CS505MU and lens MVL7000 from Thorlabs, this results in the smallest resolvable feature being 40\,$\upmu$m.
The device was mounted on a tripod in front of a view port, which allowed to capture the beam shape. 
A MagneToF detector by ETP ion detect was used for two purposes: first, to determine the time of flight (ToF) of the antiprotons (<1.5\,ns multiple ion pulse width), and second, in combination with an ``energy grid'', to determine the kinetic energy distribution of the decelerated antiprotons. 
The energy grid consists of a stack of three grids by ETP ion detect with a diameter of 76.2\,mm. 
The distance between the grids is 15\,mm.
The grid wires have a diameter of 0.018\,mm, a centre-to-centre distance of 0.25\,mm, and a transmission of 92\% to 95\%.
The two outer grids were grounded, while a blocking voltage was applied to the middle one, with a ripple of less than 10\,mV. 
The energy grids and the MagneToF detector can be moved out of the beam axis independently. 
In addition to those detectors, the BTV further upstream in the beamline (see Fig.\,\ref{Fig:setup}) was used for particle detection and intensity determination.

\subsection{Pulsed Drift Tube Switching Delay}\label{subsec:switch_delay}
\begin{figure}[!h]
	\centering
    \begin{subfigure}{\linewidth}
    \centering
        \includegraphics[width = \linewidth]{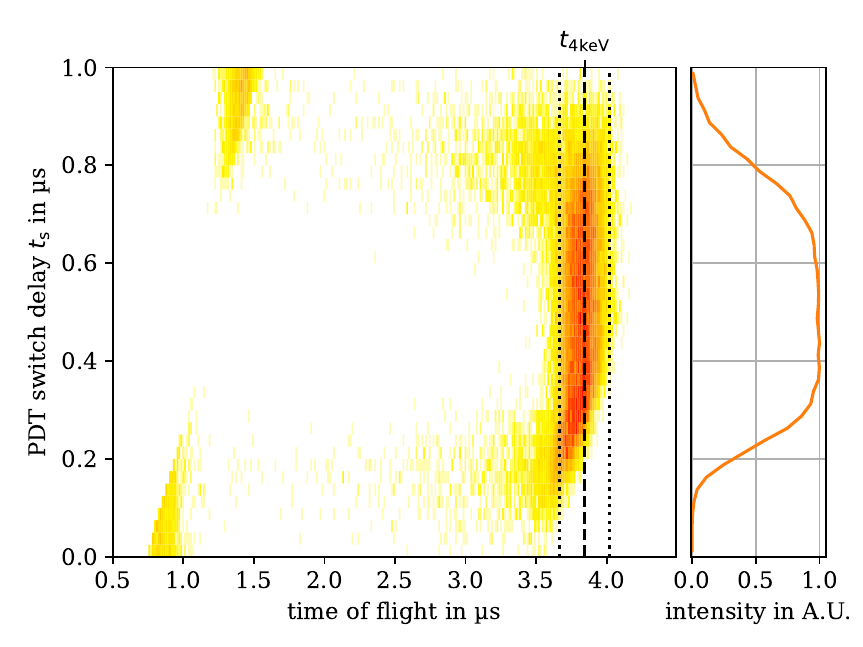}
    \end{subfigure}
    \begin{subfigure}{\linewidth}
    \centering
        \includegraphics[width = \linewidth]{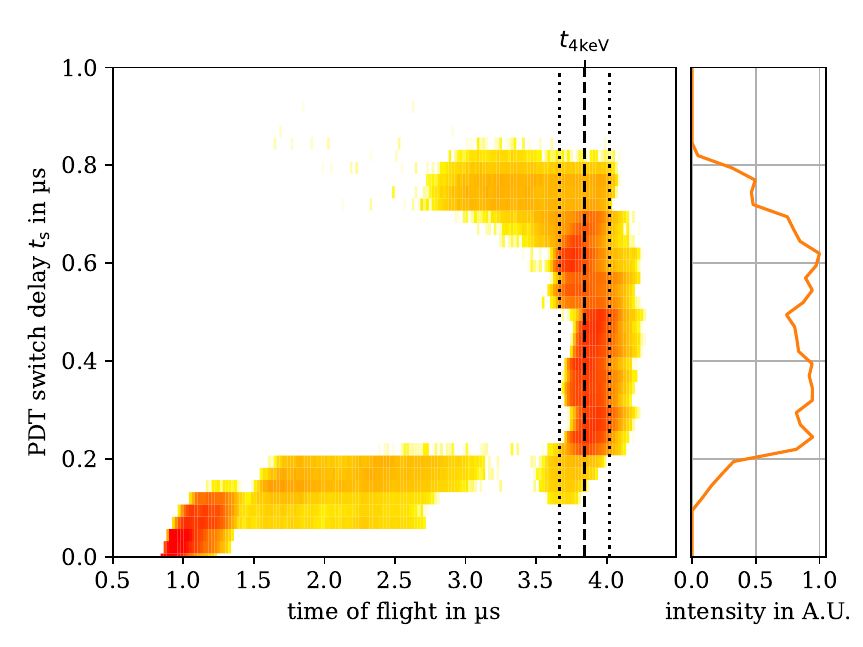}
        \end{subfigure}
    \caption{Simulated (top) and measured (bottom) beam intensity when switching the pulsed drift tube from $-96\,\si{kV}$ to ground and varying the switch delay $t_\mathrm{s}$. Yellow colours indicate lower and red higher intensity. In both cases, a successful deceleration to $4\,\si{keV}$ corresponds to a time of flight of $t_\mathrm{4keV} = 3.85\,\si{\upmu s}$, with a bunch length ($1\sigma$) of $0.09\,\si{\upmu s}$. On the right, the integrated intensity from $t_\mathrm{4keV}-2\sigma$ to $t_\mathrm{4keV}+2\sigma$ is shown, $t_\mathrm{s}$ is chosen to maximise this intensity.}\label{fig:switch-delay}
\end{figure}

When antiprotons arrive in the experimental zone, a trigger signal from the ejection from ELENA is forwarded to the electronics. 
Relative to the trigger, a switching time $t_\mathrm{s}$ has to be determined, at which the bunch is fully contained inside the pulsed drift tube, so that the deceleration is successful for the full antiproton bunch. 
To determine the ideal value, $t_\mathrm{s}$ has to be scanned while observing the time of flight of the antiprotons. 
If $t_\mathrm{s}$ is too small, the antiprotons see a grounded electrode and traverse the pulsed drift tube at full speed, arriving the earliest and with their initial energy. 
If $t_\mathrm{s}$ is too large, the antiprotons are decelerated while entering the pulsed drift tube and reaccelerated when leaving it, thus they arrive later than the ones never decelerated, but still with their initial energy. 
When switching at the correct time, the antiproton bunch is decelerated on entry but is not reaccelerated on exit. 
Thus, it arrives later than in the other cases, as they are slower, which can be seen in a simulation of the deceleration in the pulsed drift tube performed in SIMION\textsuperscript{\textregistered} (see top panel of Fig.\,\ref{fig:switch-delay}).\\
The results from the measurement can be seen in the bottom panel of Fig.\,\ref{fig:switch-delay}, they match the behaviour expected from simulations. 
When $t_\mathrm{s}$ is too small, the antiprotons arrived early. When increasing $t_\mathrm{s}$, the bunch diffuses, as the bunch is partly in the fringe field of the electrode when the pulsed drift tube is switched. 
Afterwards, in a window of about 300\,ns, the antiprotons are uniformly decelerated.
As $t_\mathrm{s}$ is further increasing, the bunch diffuses again, because it is only partly inside the pulsed drift tube when switching.\\

The measurement shows a successful deceleration, and an estimation with the time of flight gives a deceleration to $(4.0\pm 0.5)\,\si{keV}$.
A more precise measurement of the energy distribution was done using the energy grids (see Sec.\,\ref{subsec:e-spread}).

\subsection{Transmission and Focusing}\label{sec:focusing}
\begin{figure}[!h]
	\centering
	\includegraphics[width = \linewidth]{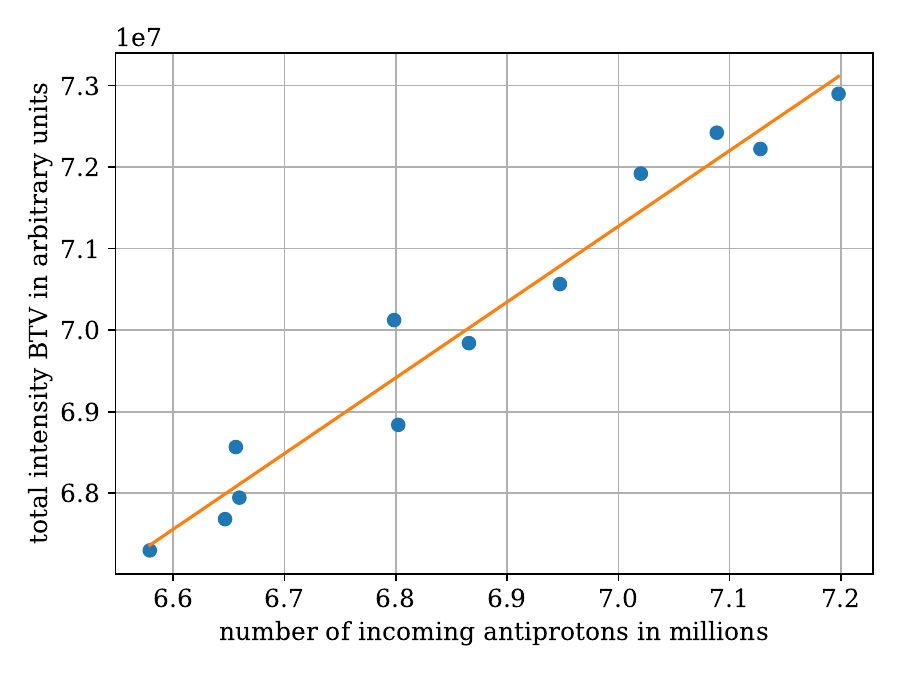}
	\caption{The bunch intensity of antiprotons determined by the ELENA detectors is proportional to total intensity on the BTV. The transmission to the BTV is $100\%$ when not decelerating the antiprotons. This allows to make a calibration to determine the transmission through the pulsed drift tube while decelerating.}
	\label{Fig:calibration}
\end{figure}

\begin{figure}[!h]
    \centering
    \includegraphics[width=\linewidth]{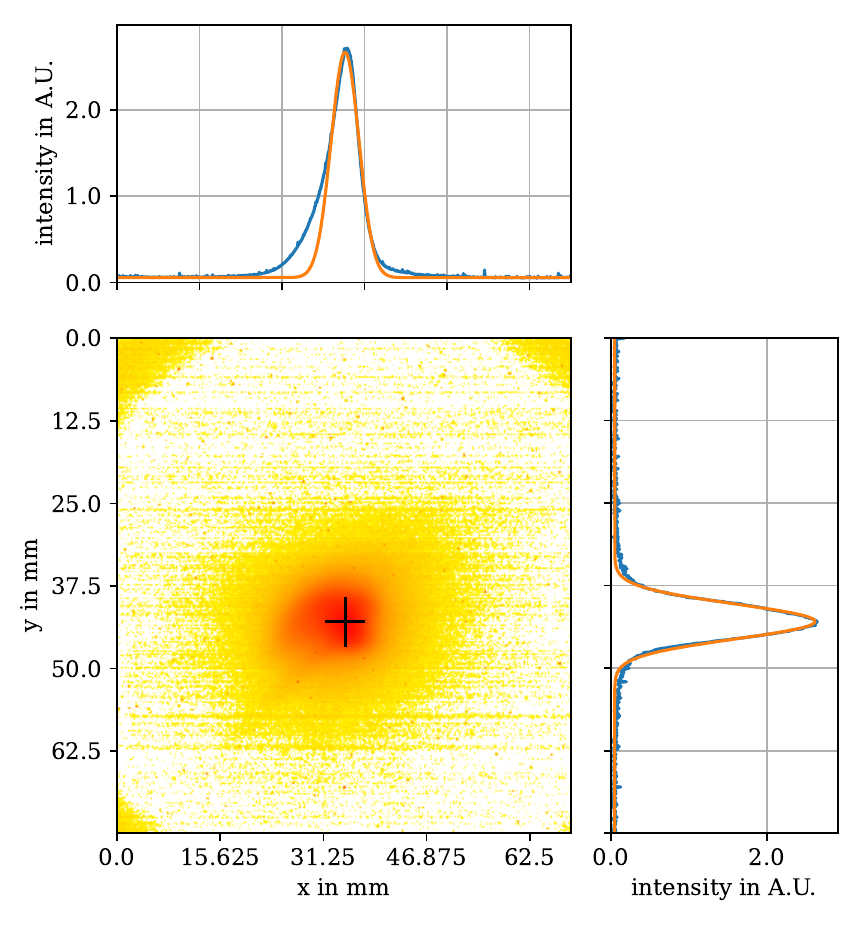}
    \caption{Beam profile after optimizing the LV einzel lenses for deceleration to $\SI{4}{\kilo\electronvolt}$ and focus on the BTV. Fitting a Gaussian to the centre peak yields $\sigma_\mathrm{horiz}=3.0\,\si{mm}, \sigma_\mathrm{vert}=3.8\,\si{mm}$. Yellow indicates a lower and red a higher intensity.}\label{Fig:focussing}    
\end{figure}

The intensity of the bunch after the pulsed drift tube $I$, can be compared to the initial intensity of the bunch $I_0$. 
The total transmission through the pulsed drift tube is thus defined by $T = I / I_0$. $I_0$ is determined before the handover point by pick-ups in the ELENA transfer lines \cite{Bartmann2018}. 
Besides showing the beam spot shape, the total intensity on the BTV is proportional to $I$, as can be seen in Fig. \ref{Fig:calibration}. 
Using the calibration in this plot, $T$ can be calculated. \\ 
$T$ for 100\,keV bunches is about 100\%. 
During the experiment, the transmission of antiprotons decelerated to 4\,keV reached ($55\pm 3$)\%, while in simulations a transmission of 100\% could be reached.
The main source of losses in transmission can be assigned to a misalignment of the high-voltage einzel lens and the pulsed drift tube and a high leakage current on the high-voltage einzel lens, which limited the voltage to -85\,kV. 
In addition, the parameters assumed in the simulation for the incoming beam might also play a role. 
Figure\,\ref{Fig:focussing} shows the beam profiles recorded by the BTV directly after the last einzel lens. 
Using a Gaussian fit, the following parameters could be obtained:
\begin{align*}
    \sigma_\mathrm{horiz}=(3.0 \pm 0.1)\,\si{mm},\,\sigma_\mathrm{vert}=(3.8\pm 0.2)\,\si{mm}
\end{align*}
64\% of the antiprotons are within a circle of radius $r=5.6\,\si{mm}$, the smallest aperture of the PUMA Penning trap. 
The focal point will have to be optimized at a later point for the injection into the PUMA trap.
\noindent

\subsection{Energy Distribution}\label{subsec:e-spread}
\begin{figure}[ht]
	\centering
	\includegraphics[width = \linewidth]{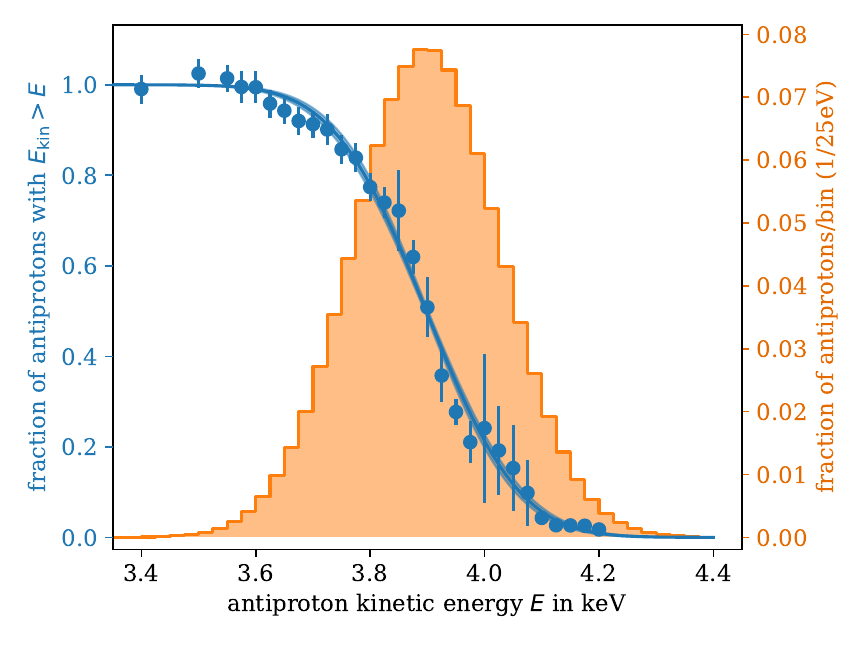}
	\caption{The energy distribution of decelerated antiprotons. The data and fitted CDF of a normal distribution are shown in blue, and the probability density function corresponding to the fit in orange. The mean energy is $\mu = (3898 \pm 3)$\,\si{eV} and the standard deviation $\sigma = 127 \pm 4$\,\si{eV}. $88\%$ of decelerated antiprotons are within $\pm 200\,\si{eV}$ of the mean energy, which is the estimated energy acceptance for trapping.}
	\label{Fig:energy-scan}
\end{figure}

The standard deviation of the ions' energy after deceleration to 4\,keV at the position of the MagneToF detector was simulated to be 101\,eV. 
The kinetic energy $E$ of the antiprotons was determined by blocking the antiprotons with the energy grids, and measuring the transmission on the MagneToF. 
The results from this can be seen in Fig.\,\ref{Fig:energy-scan}. 
In blue, the transmission onto the MagneToF is displayed in dependence of the kinetic energy of the antiprotons. 
Fitting the cumulative distribution function (CDF) of a normal distribution yields the mean energy $\mu = (3898 \pm 3)$\,eV and energy spread $\sigma = (127 \pm 4)$\,eV.
The energy distribution calculated from the fit is shown in orange.
88\% of decelerated antiprotons are within $\pm$\,200\,eV of the central energy, which is the energy acceptance for successful trapping in the PUMA Penning trap, according to simulations. 

\subsection{Bunch Length}
The length of the antiproton bunch at 4\,keV is relevant, because it determines the losses in the second stage of deceleration to a few 100\,eV right in front of the trap. 
The simulation predicts an increase in length from 75\,ns to 89\,ns at the position of the MagnetToF, with which 90\% of the bunch can be trapped.
A measurement of the bunch length of the decelerated antiprotons with the MagneToF yields a length ($1 \sigma$) of 93$\pm$3\,ns, consistent with the simulation. 

\section{Conclusion}
An overview of the design and the characterisation of the low-energy antiproton beam line of PUMA at ELENA is presented.
Design considerations for high voltage and ultra-high vacuum are discussed, as well as procedures for high-voltage conditioning and in-vacuum high-voltage operation.
The antiproton beamline is shown to be successful in decelerating antiprotons from 100\,keV to $(3898 \pm 3)$\,eV, the first step in trapping antiprotons for the PUMA experiment. 
The pressure, with the pulsed drift tube not in operation, is below $2\cdot 10^{-10}$\,mbar. 
With the implemented high-voltage ramping scheme, the pressure stays below $2\cdot 10^{-10}$\,mbar 75\% of the cycle time, also during operation.
Currently, a transmission of ($55\pm 3$)\% for antiprotons decelerated to 4\,keV can be reached.
The beam was focussed to a spot with $ \sigma_\mathrm{horiz}=(3.0 \pm 0.1)\,\si{mm},$ and $\sigma_\mathrm{vert}=(3.8\pm 0.2)\,\si{mm}$, demonstrating it can be focussed into the PUMA Penning trap.
The length of the 4\,keV antiproton bunch, relevant for the second deceleration from 4\,keV to 100\,eV is (93$\pm$3)\,ns.
Further improvement of the beamline is foreseen in the future, while the current performance already allows for first experiments with PUMA.

\section*{Acknowledgements}
We thank the ELENA team and the operators of the Antimatter Factory for excellent beam during the runs in 2022 and 2023.
We thank the technical teams at CERN and TU Darmstadt for their support.
The presented work benefited from the support of the GBAR collaboration. 
PUMA is funded by the European Research Council through the ERC grant PUMA-726276 and the Alexander-von-Humboldt foundation. 
The development of PUMA and its implementation at CERN are supported by the TU Darmstadt and CERN.

\printbibliography 
\end{document}